\documentclass{PoS}
\usepackage[authoryear,square]{natbib}
\bibpunct{(}{)}{;}{a}{}{,}

\title{The physics of Reionization: processes relevant for SKA observations.}

\ShortTitle{Physics of Reionization}

\author{\speaker{Beno\^{\i}t Semelin}\thanks{A footnote may follow.}\\
	LERMA, Observatoire de Paris, UPMC, France\\
        E-mail: \email{benoit.semelin@obspm.fr}}

\author{Ilian Iliev\\
 Astronomy Centre, Department of Physics \& Astronomy, Pevensey II
Building, University of Sussex, Falmer, Brighton BN1 9QH, United Kingdom\\
        E-mail: \email{I.T.Iliev@sussex.ac.uk}}

\abstract{
The local intensity of the $21$ cm signal emitted during the Epoch of Reionization that will be mapped by the SKA is modulated by the amount of neutral hydrogen.
Consequently, understanding the process of reionization of the intergalactic medium (IGM) is crucial for predicting and interpreting the upcoming observations. After presenting the basic physics and most meaningful  quantities pertaining to the process of reionization, we will review recent progress in our understanding of the
production and escape of ionizing photons in primordial galaxies and of their absorption in the IGM especially in so-called minihalos and Lyman Limit Systems.}

\FullConference{
Advancing Astrophysics with the Square Kilometre Array\\
June 8-13, 2014\\
Giardini Naxos, Italy}

\usepackage{amsmath}
\newcommand{\skipthis}[1]{}

\begin{document}

\section{Introduction}`

The SKA will observe the Epoch of Reionization (EoR) through the redshifted $21$ cm emission of hydrogen. The intensity of the emitted signal depends on several local quantities but during most of the EoR the most important one is obviously the ionization state of the gas: only neutral atoms in the ground state can emit $21$ cm radiations. Fully ionized regions will show up as null-signal areas in the tomographic data cube. While the basic physics of hydrogen ionization is simple, it takes place in a strongly inhomogeneous medium and thus involves vastly different scales and characteristic times, resulting in a complex geometry and evolution of the ionization field.

To discuss the important aspects of the reionization process we will first consider that the gas consists of pure hydrogen. The effects of the  presence of helium and metals will be discussed as needed. If $x_i$ is the local ionization fraction of hydrogen, it
obeys the simple equation:

\begin{equation}
{d x_i \over dt} = \gamma(T) (1-x_i) n_e - \alpha(T) x_i n_e + \Gamma (1-x_i) \quad\quad \mathrm{with} \quad\quad \Gamma=\int d\Omega \int_{\nu_0}^\infty {\sigma_\nu I_\nu \over h \nu} d\nu
\end{equation}
\noindent
where $n_e$ is the number density of electrons, $T$ is the temperature, $\gamma$ is the collisional ionization coefficient, $\alpha$ is the recombination coefficient, $\Gamma$ is the photoionization rate, $\nu$ is the frequency, $I_\nu$ is the local specific intensity, $\sigma_\nu$ is the ionization cross-section, and $\Omega$ the solid angle. 
Tightly coupled to the ionization equation is
the energy equation:

\begin{equation}
{dE \over dt} = \mathcal{H}- \Lambda
\end{equation} 
\noindent
where $E$ is the internal energy of the gas, $\mathcal{H}$ is the heating rate and $\Lambda$ is the cooling rate. Both $\mathcal{H}$ and $\Lambda$ include contributions from radiative processes (photo-heating, collisional ionization cooling, recombination cooling, collisional excitation cooling, Bremsstrahlung cooling and Compton cooling - heating) and from dynamical processes (adiabatic cooling or heating, heating from viscous dissipation). References for fitting functions of the
various rates, that depend on temperature and ionization state and for the ionization cross-section that depends on the frequency can be found for example in \citet{Iliev06b}.

\section{Insights into the ionization process}

Along with the equations above, the process of reionization is governed by the radiative transfer equation \citep{Rybicki86}. In the general case, these equations are furthermore coupled
to the hydrodynamics equation of the self-gravitating gas through the energy equation and the equation of state of the gas. However ionization by UV photons will only heat up the gas to $\sim 20\,000$ K, and the amount of X-rays produced during the EoR will not change this
picture in the bulk of the IGM. The associated increase in the pressure of the gas will affect noticeably only structures with a virial temperature of the same order as the heated gas temperature or lower. If the radial density profile of real halos actually follow the numerical fit in  \citet{Navarro97} (the so-called NFW profile), structures with a virial temperature less than $\sim 20\,000$ K have a total masse $\lesssim 10^8 M_\odot$ (both dark matter and baryons included). The gas in these structures may  be photo-evaporated by the arrival of an ionization front or will never fall into the potential well of halos in ionized regions.
On larger scales though, the coupling to the dynamics may safely be ignored as gravitation dominates over hydrodynamic pressure.

Evaluating a few physical quantities in typical EoR environments will help us get a first idea on
how reionization proceeds.

\subsection{Mean free path of ionizing photons}

In a homogeneous medium the mean free path of ionizing photons is defined by 
$l={1 \over \sigma_\nu n_H}$, where $\sigma_\nu \simeq 6.3\, \, 10^{-18} \left( {E \over E_0} \right)^{-3} \,\, \mathrm{cm}^{-2}$ is the ionization cross-section (see \citet{Verner96} for a more accurate formula), $E_0$ is the energy of the ionization threshold and $n_H$ the number density of absorbers, here neutral hydrogen atoms. In a fully neutral IGM at the average
density of the universe we find $l \simeq 2. \, \left({10 \over 1+z}\right)^2 \left({E \over E_0}\right)^3$ comoving kpc. In higher density regions the value will be even smaller. Consequently,
a reionization fueled by stars that produce ionizing photons at energies not much higher than the ionization threshold will proceed in the form of fully ionized bubbles growing around the sources, separated from the neutral region by a sharp ionization front. On the other hand, the small amount of 
X-ray photons emitted above
1 keV have a mean free path in the 1 comoving Gpc range and will provide a weak (a few percent at most) and mostly uniform contribution to the ionization field (note that the associated heating has important consequences for the $21$ cm signal). Photons with energies of a few keV have only a small probability of being absorbed in the IGM before reionization is complete.

\subsection{Recombination time}

Another important quantity is the typical recombination time. Or, in other words, how many times does
an hydrogen atom in the IGM need to be ionized by the end of the EoR. The recombination time is 
defined as $t_\mathrm{rec}={1 \over \alpha(T) n_H}$. At the average density of the universe and for a gas at $T=10^4$K, we find, $t_\mathrm{rec} \simeq 240 \left( {10 \over 1+z} \right)^3 $Myr. Since the bulk of the IGM is likely reionized in the $6<z<10$ range, most atoms need to be ionized only once. At the same
time, a moderate overdensity of ${\delta \rho \over \rho} \sim 20$ will have a much shorter recombination time of $\sim 10$ Myr. Such structures (filaments, minihalos) will act as photon sinks
until the ionizing flux is large enough to photo-evaporate them \citep[see, e. g.][] {Sobacchi14}. The recombination time in the interstellar medium will be even shorter and, most likely, only the clumpiness of the medium or holes blown
by supernovae explosion or AGN feedback will allow ionizing photons to escape into the IGM (see next section).

\subsection{Ideal Stromgren sphere model}

Assuming sharp ionization fronts, neglecting recombinations in ionized regions and neglecting the coupling to the hydrodynamics are thus reasonable assumptions for building
simple analytic models on how the IGM is reionized. If we further simplify by assuming an isolated point source in a homogeneous IGM and a constant temperature in the ionized region we get the well known Str{\"o}mgren sphere model \citep{Stromgren39}. In this case the radius of the spherical ionized region grows as: $r(t)=r_s \left[1-\exp(t/t_{\mathrm{rec}})\right]^{1/3}$ where $r_s=\left[ {3 \dot{N}_\gamma \over 4 \pi\alpha(T) n_H^2 }\right]^{1/3}$ is the asymptotic radius of the ionized sphere and $\dot{N}_\gamma$ is the number of ionizing photons produced per second by the source. From this model we can estimate  that the typical velocity of ionization fronts in the IGM during the EoR is $\gtrsim 10^3$ km.s$^{-1}$. Let us emphasize that during the EoR, the luminosity of proto-galaxies typically increases on time scales much shorter that $t_{\mathrm{rec}}$, and thus the asymptotic Str{\"o}mgren radius is never reached. 

Generalized solutions for the Str{\"o}mgren sphere in an expanding universe 
have been formulated \citep{Shapiro87}. In view of the
future attempts to detect individual bubbles in the SKA tomographic data, it is also important to notice that the apparent
shape of the bubbles along the light cone is not isotropic \citep{Yu05,Majumdar11}.

\section{Ionization processes in primordial galaxies}

\subsection{Escape of the ionizing continuum}
 The ionization history of the IGM is regulated by the number of ionizing photons escaping from primordial galaxies. From an observational point of view this number can be inferred from the galaxy luminosity function at various redshifts. The most recent data give us indications to the amplitude and shape of the luminosity function up to $z=8$ \citep[e.g.][]{Bouwens14}. But then modeling comes in to connect the luminosity function to the mass function, through a $M/L$ ratio that
 may depend on a number of parameters. For an in-depth understanding of the physics of primordial galaxies however, it is useful to break the $M/L$ parameter in two parts: the star formation rate (SFR) and the radiation escape fraction $f_\mathrm{esc}$, and more specifically  for our purpose
 the escape fraction of the ionizing continuum. Indeed, primordial galaxies host dense gas clouds with short
 recombination time that will consume a large fraction of the emitted photons before they ever reach
 the slowly-recombining diffuse IGM. Dust may also play a role in absorbing photons during the later phase of the EoR. In any case, $f_\mathrm{esc}$ and its dependence on the host galaxy and its environment is the result of complex radiative processes.
 
From an observational point of view, measuring $f_\mathrm{esc}$ during the EoR presents a challenge
since the photons that do escape galaxies will be absorbed in the IGM and participate in the global reionization process, and thus will never reach us. Indirect probes can be used, such as the determination of the average photoionization rate in the IGM from Ly-$\alpha$ forest observations, or
the Lyman-$\alpha$ emitters (LAE) luminosity since the intensity of the Lyman-$\alpha$ line is proportional to the recombination rate in the galaxy and thus to the amount of ionizing radiation that does not escape. Using such observations, a substantial level of modeling is needed to yield $f_\mathrm{esc}$.
 Observations provide some constraints for the relative escape fraction between ionizing and non-ionizing photons  at $z \sim 3$ \citep[e.g.][]{Steidel01, Iwata09}. A few studies focused on higher redshift provide only tentative results: an upper
limit of 0.6 at z=5.6 \citep{Ono10} or an inferred value of 0.3 at z=6-8 \citep{Finkelstein12}. However, there is a general agreement based on models that, if
stellar sources are to be the main contributor to reionization, the escape fraction must increase
toward higher redshifts and/or be higher in low mass galaxies \citep[e.g.][]{Alvarez12, Haardt12, Kuhlen12, Mitra12}. 

Another way to produce estimates of $f_\mathrm{esc}$ is  to run numerical simulation that preferably include the coupling between the ionizing continuum radiative transfer and the dynamics. These simulations need to resolve to some extent the interstellar medium while preserving the cosmological environment and have become possible only recently \citep{Gnedin08, Wise09, Razoumov10, Yajima11, Yajima14, Paardekooper13, Wise14, Kimm14}. They do not use similar resolution nor similar modeling, and thus do not quantitatively converge in terms of the mean value and scatter of $f_\mathrm{esc}$ as a function of redshift and host halo mass. One common trend seems the emerge however: $f_\mathrm{esc}$ decreases with the mass of the host halo (with \citet{Gnedin08} finding the opposite trend however, and \citet{Kimm14} finding almost no dependence), and increases with redshift (weak opposite tend in \citet{Kimm14}). This seems to indicate that low mass galaxies  made the largest contribution to the production of ionizing photons, at least at high redshift, before radiative feedback quenched their star formation rate.

\subsection{Feedback on primordial galaxies}

The second important aspect of ionizing radiative processes in primordial galaxies is the feedback on the gas (and thus star formation). This feedback is enacted through the photo-heating and photo-evaporation of the  gas in galaxies and of the accreting gas clumps and filaments, but also to some extent through radiation pressure.

In a forming galaxy made of near-primordial gas, cooling processes  below $10^4$ K are rather inefficient (low metal content, uncertain H$_2$ formation). Thus photo-heating the cold gas back to $\sim 10^4$ K through ionization may substantially  decrease the SFR. This is easily understood in $< 10^8$ M$_\odot$ galaxies that have a virial temperature below $10^4$ K: the gas will be expelled from the galaxy \citep{Petkova11, Dale12, Hasegawa13}. But it may also be true in more massive galaxies where the gas is not expelled but merely prevented from fragmenting. Such an effect is found by Hasegawa \& Semelin (2013) for example. Such small scale processes can be implemented in large scale reionization simulations using sub-grid recipes \citep{Iliev07}

On top of the feedback mediated by thermal pressure, radiation pressure may play a role in regulating star formation in primordial galaxies. \citet{Wise12} observed, using numerical simulation, a 
significant quenching of the star formation rate when ionizing radiation pressure is included. This result has not been reproduced yet by other
teams using different numerical techniques. It is definitely worth further investigation.

Last but not least, the feedback from supernovae explosions definitely regulates the escape fraction. Although not a pure radiative process, it shapes the gas distribution around the sources and influences the escape fraction of ionizing photons \citep{Kimm14}. 

The relative importance of these different processes probably depends on the mass of the host halo and is not the object of a consensus. Further numerical investigation is needed.

\section{Ionization processes in the IGM}

As we have seen in section 2.1, the mean free path of ionizing photons in typical IGM conditions during the EoR is very  small at the ionization energy threshold compared to typical cosmological distances. Consequently the ionization process sourced by UV photons takes the form of ionizing fronts sweeping through the IGM. The isothermal sound speed in the ionized gas behind the front is $\sim 10$ km.s$^{-1}$, and much lower in the neutral gas ahead of the front. The ionization front velocity (set by the local ionizing flux and baryon density alone if recombination is neglected) is typically larger than
$1000$ km.s$^{-1}$ in the IGM at $z\sim 10$ (see section 2.3). Consequently the gas is unable to respond dynamically, and the front is an extreme weak-R type (R for rarefied, see Spitzer 1978), with negligible compression of the gas behind the front.
This is the main justification for not coupling radiative transfer to hydrodynamics in many EoR simulation. It is worth mentioning however, that the ionization front not only slows down but even stalls in structures with moderate overdensity ($\sim 30$) due to recombinations (e. g. \citet{Sobacchi14} and references therein).

While there is little doubt that stars contributed a significant faction of the ionizing photons through their UV continuum, we have few constraints yet on the contribution of X-ray sources (X-ray binaries, Supernovas, AGN and more) to the reionization process. Because of their long mean free path, a reionization powered by X-rays only would show a diffuse and fluctuating ionization field without any fronts. It is however unlikely that X-rays contribute as much as the UV continuum to the ionization budget. Consequently the most likely picture is that
of an IGM with fully ionized regions around the sources, separated by a sharp ionization front from near-neutral regions, ionized at most a few percent  by X-rays. An important effect of X-ray penetrating into the neutral IGM however is the heating by secondary
electrons \citep{Furlanetto06,Shull85}. Indeed when an X-ray photon ionizes an hydrogen atom, the energy in excess of the ionization threshold goes into kinetic energy for the freed electron. This electron then collides with neighboring atoms, producing secondary ionizations, excitations and heating. Heating up the cold IGM by a few K only has a strong impact on the $21$ cm signal.

The presence of helium in the primordial gas has a moderate impact on the reionization
of hydrogen. With first and second ionization energies at $24.6$ eV and $54.4$ eV respectively, Helium filters and hardens the spectrum of Pop III and Pop II stars. In practice the emissivity above $54.4$ eV is small enough that the reionization of HeII is negligible at $z>6$. On the other hand, HeI is reionized mostly at the same time as HI, moderately slowing the pace of the process by consuming a small part of the radiations \citep[e.g.][]{Ciardi12}. Behind the ionization fronts, the temperature of the ionized IGM is somewhat higher than if Helium is not considered.
Overall, the impact of helium reionization on the $21$ cm signal is small.

\section{Recombination processes in the IGM}
\subsection{Minihalos}
The very first stars formed in cosmological minihalos, much smaller than 
present-day galaxies, with total masses below $\sim10^8M_\odot$. The efficiency of this process is however uncertain. Indeed, the Jeans mass
criterion for star formation requires that the gas be extremely cold, with 
temperatures of just a few hundred K. At primordial composition, consisting 
almost exclusively of Hydrogen and Helium, the gas in such small halos can 
only cool through $H_2$ molecular line cooling. These molecules are fragile, however, 
and easily dissociated by Lyman-Werner radiation. Moreover, minihalos are photoevaporated
by ionizing radiation \citep{2004MNRAS.348..753S,2005MNRAS.361..405I},
which is a complex process that results in screening of the intergalactic
medium by them and slowing down the large-scale ionization fronts 
\citep{2005ApJ...624..491I,2006MNRAS.366..689C}.

\begin{figure}[t]
\centering
\resizebox{!}{6.5 cm}{\includegraphics{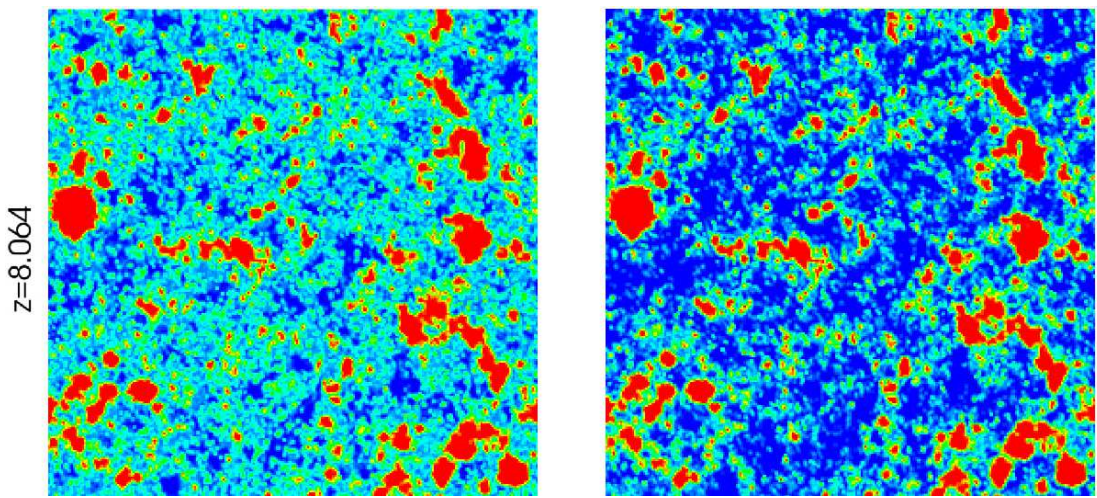}}
\caption{Maps of the ionization fraction color coded form 0. (blue) to 1. (red) in simulations modeling the effects of minihalos (left) or not (right). From \cite{2012ApJ...756L..16A}.}
\label{MHs_vs_noMHs}
\end{figure}

Furthermore, due to their shallow gravitational potential wells, the minihalo 
abundance is influenced by heating of the IGM by e.g. X-rays, as well as 
modulated locally by baryon-dark matter velocity offsets \citep{Tseliakhovich10}. 
The relative importance of these effects is still highly uncertain and requires 
much better knowledge than we currently possess of the early sources of X-rays and soft UV radiation: their 
nature, abundance and clustering. Nonetheless different 
aspects of this physics have been modeled in both simulations 
\citep{2009ApJ...695.1430A,2010A&A...523A...4B,2012ApJ...756L..16A} and semi-analytical 
work \citep{2010MNRAS.406.2421S}. Both X-rays and Lyman-Werner radiations have very long mean
free paths, of order 100 Mpc or more, and the baryon-dark matter velocity 
modulation acts on similarly large scales, all of which could result in significant 
fluctuations in the local abundance of minihalos and their star formation rates 
\citep[e.g.][]{Fialkov13}. The first stars forming in cosmological minihalos 
likely dominated the early reionization, but eventually the Lyman-Werner radiation from those 
same stars suppresses further star formation, stalling the reionization history until
the larger-mass, atomically-cooling halos take over \citep{2012ApJ...756L..16A}. This 
results in significantly different early reionization history compared to the 
case without minihalos, much higher electron-scattering optical depth for CMB photons and a different reionization geometry
(see Figure \ref{MHs_vs_noMHs}).

\subsection{Lyman Limit systems}

Lyman Limit Systems (LLS) are structures of any mass that are optically thick to the UV background (in practice with column density larger than $10^{17}$ cm$^{-3}$). They act as screens for UV photons traveling in ionized regions.
Analysis of QSO spectra show that LLS limit the mean free path of
UV photons to $\sim 50$ cMpc at $z=6$ \citep{Songaila10}. Estimates at higher redshifts are uncertain. The main consequence
of LLS would then be to slow down the end of reionization (in terms
of the average ionization fraction reaching $1$), by limiting the
effective number of sources that contribute to the local ionizing flux.

LLS are difficult to take into account in full numerical simulations because it typically involves resolving the internal structure and internal ionization processes in dwarf galaxies. Subgrid models can be devised and implemented. Using semi-numerical simulations \citet{Crociani11} and \citet{Sobacchi14} find that LLS have an impact on the history and geometry of reionization and thus on the $21$ cm signal power spectrum. \citet{Alvarez12b} also implement the effect of LSS in simulations.

\section{Conclusion}

At first glance, the process of reionization is simple: ionized bubbles around primordial galaxies with sharp, supersonic ionization fronts expanding into a neutral IGM mildly heated and weakly ionized by X-rays. It is however regulated by two complex processes: the production and escape of ionizing photons from primordial galaxies, and the presence of dense, small scale structures in the IGM that act as photon sinks (minihalos and LLS). Both of these processes have an impact on the pace and geometry of reionization. On the one hand observations provide limited constraints on these processes at $z<6$ whose extrapolation during the Eopch of Reionization is risky. On the other hand, simulations do not offer a consensus on these processes, mainly because they involve small scales that are difficult to resolve while keeping a statistically significant volume. Either significant progress will be made in the next few years or the interpretation of the observations of the $21$ cm signal with the SKA will have to deal with models
involving a large parameter space. 

\bibliographystyle{apj}
\bibliography{myref,references}








\end{document}